# Exploratory and Live, Programming and Coding

## A Literature Study Comparing Perspectives on Liveness

Patrick Rein[a], Stefan Ramson[a], Jens Lincke[a], Robert Hirschfeld[a], and Tobias Pape[a]

a    Hasso Plattner Institute, University of Potsdam, Germany

**Abstract**    Various programming tools, languages, and environments give programmers the *impression of changing a program while it is running*. This experience of *liveness* has been discussed for over two decades and a broad spectrum of research on this topic exists. Amongst others, this work has been carried out in the communities around three major ideas which incorporate liveness as an important aspect: *live programming*, *exploratory programming*, and *live coding*.

While there have been publications on the focus of each particular community, the overall spectrum of liveness across these three communities has not been investigated yet. Thus, we want to delineate the variety of research on liveness. At the same time, we want to investigate overlaps and differences in the values and contributions between the three communities.

Therefore, we conducted a literature study with a sample of 212 publications on the terms retrieved from three major indexing services. On this sample, we conducted a thematic analysis regarding the following aspects: motivation for liveness, application domains, intended outcomes of running a system, and types of contributions. We also gathered bibliographic information such as related keywords and prominent publications.

Besides other characteristics the results show that the field of *exploratory programming* is mostly about technical designs and empirical studies on tools for general-purpose programming. In contrast, publications on *live coding* have the most variety in their motivations and methodologies with a majority being empirical studies with users. As expected, most publications on live coding are applied to performance art. Finally, research on *live programming* is mostly motivated by making programming more accessible and easier to understand, evaluating their tool designs through empirical studies with users.

In delineating the spectrum of work on liveness, we hope to make the individual communities more aware of the work of the others. Further, by giving an overview of the values and methods of the individual communities, we hope to provide researchers new to the field of *liveness* with an initial overview.

# The Art, Science, and Engineering of Programming

**Exploratory and Live, Programming and Coding**

# 1 Introduction

A variety of programming environments and tools can provide the *impression of changing a program while it is running* [8, 36, 38]. Nowadays, this impression is often described as *liveness*. While research on *liveness* is not part of mainstream research on programming, still a broad spectrum of contributions exists. In particular, three ideas incorporate this capability as an integral part: *live coding, live programming*, and *exploratory programming*. As each term has its own research community, shared approaches and potential synergies might be overlooked. Further, the whole range of research on liveness only becomes visible when looking at all of these communities. To delineate this potential across communities and illustrate the range of contributions, we conducted a literature study.

Looking at prominent publications from each community, general differences become apparent. Generally speaking, *live coding* is often concerned with the creation of art through changing source code as a performance in front of an audience [1, 8, 37]. *Live programming* in contrast often seems to put the very activity of programming in its focus [13, 35]. Correspondingly, the term seems to be used when describing programming tools which provide immediate feedback on the dynamic behavior of a program even while programming. The term *exploratory programming* often refers to a particular workflow during programming whenever requirements are not fully defined but are yet to be discovered [31, 38]. It is supported by exploratory programming environments that incorporate changing a running system to make exploration of unknown domains or of design alternatives easier. Notably, the term also often refers to programming practices which do not require liveness. For this study, however, we refer to the usage of this term in which liveness is relevant [31, 38].

The communities around these ideas differ in various regards such as their motives for dealing with changing a program while it is running, the fundamental perspective on the activity of programming, applications domains in which liveness is used, and desired outcomes of their research (system architectures, workflows, experiences). Besides their differences, they share the common notion of creating an impression of changing a program while it is running. Their differing approaches to this notion could be leveraged for cross-pollination between the communities to advance all three of them. For example, *exploratory programming* systems could benefit from the specialized mechanisms and tools investigated under the term *live programming*. At the same time the *live programming* tools could be examined through the lens of experience reports from programmers as it is done in the *live coding* community.

All three communities are academic communities. Thus, we treat the publications published under the corresponding terms as artifacts which can be studied to gain insights into the values and practices of each community. Hence, we relate the communities around these terms to each other by surveying the similarities and differences in a sample of the corresponding literature. In particular, our contributions in this paper are:

**Patrick Rein, Stefan Ramson, Jens Lincke, Robert Hirschfeld, and Tobias Pape**

- A systematically collected sample of 212 publications indexed in the ACM DL, IEEE Xplore, and DBLP on the terms *live programming*, *live coding*, and *exploratory programming*.[1]
- A thematic analysis [2] of the sample regarding the motivations for liveness, applications domains in which the liveness is used, types of contributions, historic distribution of terms, intended outcomes of programming, prominent publications for each term, and other fields related to the three terms
- A comparison of the results of the thematic analysis between the corpus for each of the three terms

The remainder of the paper is structured as follows. We begin with the specific research questions guiding the literature study in section 2. We will then illustrate our methodology in section 3. We describe the characteristics of the three corpora in section 4 and the results of the study and the relation to the research questions in section 5. We discuss correlations in the study results, threats to validity, and potential consequences in section 6 and summarize our findings and point out further potential research in section 7.

## 2  Research Questions

The goal of this survey is to determine differences and commonalities between the three communities around the terms *live coding*, *live programming*, and *exploratory programming*. In this context, we posed seven research questions.

1. The notion of an impression of changing a program while it is running has been around since the 1950s and the term *liveness* has been coined in the 1990s [23, 36]. However, communities around these terms seem to have formed later, thus we want to determine: *What is the distribution of publications on the different terms over time?*
2. Each community has publications which define their field. Assuming that these are the most cited works in their field we want to find out: *What are the most prominent publications?* Further assuming, that these works also shaped their respective fields we want to determine: *In which fields or intellectual contexts are the most prominent publications rooted?*
3. While live coding seems to be mainly focused on programming for performance art, live programming and exploratory programming do not seem to have such a strong focus on one application domain. Therefore, we want to find out: *In the context of which domains has liveness been used?*
4. The motivation for applying the technical concept for liveness can be telling about the values of the communities, thus we want to determine: *What are the motivations for having liveness?*

---

[1] We have published a BibTeX file of the sample and the most prominent publications. [27]

5. In order to determine what kind of knowledge each community contributes, we want to investigate: *Which kinds of contributions are described?*
6. Besides the general motivation for liveness, the general reason for modifying running systems varies greatly between different approaches. For example, the Smalltalk community regards their systems as *live* as they are constantly running and any change to a Smalltalk program is actually a change in the object graph of the running system. In order to determine whether the three communities have a particular stance on this aspect, we want to determine: *What is the general reason for having a running version of the program while changing it?*
7. The three terms used in this study are only one way of categorizing the field or research on programming. Similar and overlapping perspectives might indicate further related work and future research potential. The additional keywords used by authors for their publications can provide hints to such other related fields. Thus, we want to determine: *Which keywords have authors applied to their work?*

## 3 Methodology

In order to determine the range of topics and compare the literature related to the three terms, we conducted a thematic analysis on three corpora, one for each term. We did not aim to create a comprehensive systematic literature survey of the three terms but to determine the general distribution of themes of each corpus regarding to our research questions.

In general, we followed the SALSA (Search, AppraisaL, Synthesis, Analysis) process [12]. First, we created an overall corpus with a systematic search and an appraisal of works. We then synthesized data for the individual research questions by extracting data from the publications and by doing a thematic analysis [2]. We analyzed the resulting data by determining correlations, comparing overlaps, and differences between the corpora defined by the three terms.

As the term *coding* is relevant to the domain, we will, for the remainder of the paper, refer to the process of annotating papers with categories as *tagging*. Correspondingly, we will refer to *codes* as *tags*.

### 3.1 Choosing the Terms

In an initial search we looked at a number of terms which might yield relevant work for research on "liveness". Due to the academic venues on these terms, we chose "live programming" and "live coding".

We also wanted to cover the perspective of Lisp and Smalltalk systems but the mere names of these systems are not adequate terms as they only describe the artifacts not the particular experience of liveness inherent in these systems. The term exploratory programming subsumes these systems explicitly and describes their liveness. Other terms covering those two systems such as "self-sustaining systems" or "image-based systems" do not capture the aspect of liveness.

Other terms yielded a large portion of irrelevant or contradictory results. For example, the term "conversational programming" yields work on read-eval-print loops (REPLs) but also on computer aided design (CAD) programming which seldom includes an element of liveness. Overall we also looked at:

- immediate programming
- conversational programming
- immediate feedback in programming
- interactive programming

### 3.2 Search: Systematically Determined Corpora

For each term we created an initial corpus and then created the hull of references. We determined the initial corpora by using the full-text search on three major indexing services:

- IEEE Xplore Library (http://ieeexplore.ieee.org/search/searchresult.jsp?queryText=.QT.KEYWORD.QT.&newsearch=true)
- ACM Digital Library (DL): Guide to Computing (http://dl.acm.org/exportformats_search.cfm?query=KEYWORD&filtered=&within=owners%2Eowner%3DGUIDE&dte=&bfr=&srt=%5Fscore&expformat=bibtex)
- Digital Bibliography & Library Project (DBLP) (http://dblp.uni-trier.de/search/publ/api?q=KEYWORD%24&h=1000&format=bib1&rd=1a)

We retrieved these search results through a semi-automated process using the URLs denoted in brackets (insertion of the keyword denoted as KEYWORD). Depending on the capabilities of the search API the keywords were either entered as exact matches (for example "live programming" at IEEE) or AND queries (for example live+programming for DBLP). The retrieval from the three indexing services spanned from the 7th of July 2017 to the 22nd of September 2017.

Further, we manually added publications from manually selected scientific venues (For a detailed list of keywords and venues per term, see table 1). Namely, we added the International Conference on Live Coding (ICLC) and the Live Programming Workshop (LIVE). Other major venues, such as CHI, UIST, or PLATEAU are fully indexed by the three services we used and thus appeared automatically through our initial search. We then selected appropriate publications from this initial corpus using our appraisal criteria.

### 3.3 Appraisal

The appraisal criteria are concerned with the nature of the publication as well as the content of publications. The complete appraisal process was done manually.

For a publication to be accepted it has to be peer-reviewed. This covers publications for conferences, workshops, and journals, as well as doctoral theses. To ensure that the papers contained enough information for us to evaluate the content, we removed all publications with only one page.

**Exploratory and Live, Programming and Coding**

▮ **Table 1** Keywords and venues used to find publications for each term.

| Term | Keywords | Venues |
|---|---|---|
| live coding | livecoding, live coding | ICLC 2015, 2016, 2017[a] |
| live programming | live programming | LIVE 2013, 2016, 2017 |
| exploratory programming | exploratory programming | PX 17.2 |

[a] Only contains the publications which were actually made available from the conference website.

In general, a publication is added to a term-specific corpus if it contains the corresponding term in the title, keywords, abstract or the title of the publication it is part of, for example the title of the proceedings. Further, whenever the term occurred we manually determined whether it actually referred to the particular experience of programming we want to investigate. This is necessary as the three terms live programming, live coding, and exploratory programming are also used in other domains. Consequently, we only selected papers adhering to the following notion of programming:

> "the human activity of describing a process run by some computer"

As a consequence, we excluded publications on the following topics from the corpora:

- planning and modification of television schedules
- encoding video streams during recording and broadcasting
- algorithms for artificial intelligence which explore a solution space
- planning health programs
- analysis of concurrent programs for the liveness property

Further, the term "live coding" is used in *teaching research* to refer to a class room constellation in which teachers program "live" in the sense that students can follow the teachers' activities in a programming environment on a projector. We also excluded publications on this topic as they do not include an explicit notion of changing a program while it is running. We are aware that the writing of source code in front of an audience is an integral element of live coding. However, the fact that the mere writing of source code is done "live" in front of an audience (without continuously executing the program) does not entail any particular connection between the code and its dynamic behavior.

Also, while exploratory programming environments refers to environments which often include some form of liveness, the single term "exploratory programming" can also refer to a particular workflow [38]. This "exploratory workflow" is creating new alternatives and trying them out. This does not entail liveness but can also be done in an edit-compile-run cycle with a long round-trip time. We excluded papers exclusively referring to exploratory workflows and did not mention exploratory programming environments.

### 3.4 Synthesis: Thematic Analysis

We derived the data for research questions 1, 2, and 7 directly and manually from the content of the individual publications. The specific procedures used are reported in section 5.

As the goal of the study is to illustrate the spectrum of research on the topic of liveness, we applied thematic analysis instead of a predefined tagging scheme to gather data for research questions 3 to 5 [2]. That means we determined the tags for the themes application domain, motivation, and types of contribution during tagging. Specifically, we used theoretical thematic analysis with the themes defined by our research questions [2]. The tags for the intended outcomes of running a system (research question 6) were determined beforehand.

All publications were tagged as one corpus and in alphabetical order to prevent bias of the tagger towards tags who they might deem fitting to the corpus a publication belongs to. The tags are described in detail in the related sections in section 5. All tagging was done by one tagger. During the first pass the set of tags changed, as tags were added or the focus of a tag changed. Thus, we did a second pass through the corpus for each theme revising the tags.

Afterwards, the tagger formalized the tags in the descriptions. We determined the inter-tagger reliability for a 20 % sample using Cohen's kappa. We report the inter-tagger reliability with each result.

### 3.5 Analysis

The main goal of the study is to illustrate the spectrum of research on liveness. Thus, we directly report the determined tags for the overall corpus. We do not report on the distribution of tags for the overall corpus, as the three terms are disproportionately represented and consequently the overall distribution would be skewed to the distribution of the term with the most publications.

Another goal of the study is to investigate the overlap and differences of the terms *live programming*, *live coding*, and *exploratory programming*. Hence, as a further analysis we compare the distribution of the tags between the three term-specific corpora.

## 4 Corpus Characteristics

We can already gain some insights from the mere distribution of publications together with the bibliographic information of the publications. In this section, we will describe the distributions of publications resulting from search and appraisal. We will also illustrate the distribution of bibliographic data, such as the historic distribution of terms, related keywords, and venues.

**Exploratory and Live, Programming and Coding**

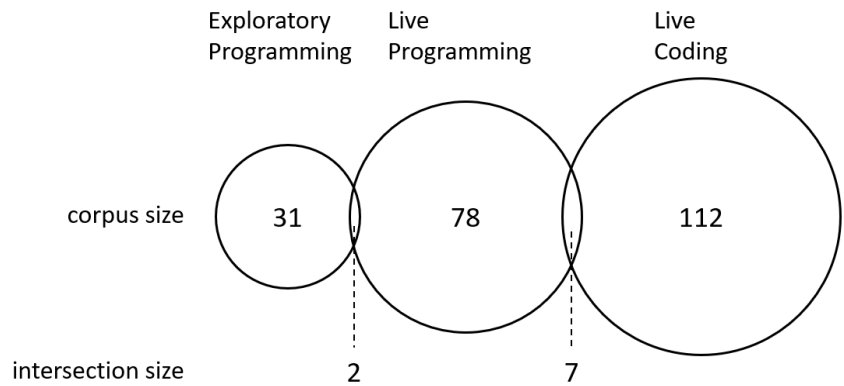

Figure 1 A Venn diagram showing the overlap between the three corpora.

## 4.1 Results of Search and Appraisal

Our search yielded 520 individual publications (see table 2). The appraisal reduced the number of publications in each corpus to 40 % to 42 %. The numbers for the individual corpora do not add up to the total value as a publication can be in more than one corpus.

Table 2 The number of publications resulting from the search phase and the appraisal phase. The percentage in brackets is the portion of publications form the search phase which were selected in the appraisal phase.

| Phase | Exploratory programming | Live programming | Live coding | All |
|---|---|---|---|---|
| Search | 77 | 187 | 269 | 520 |
| Appraisal | 31 (40 %) | 78 (42 %) | 112 (42 %) | 212 (41 %) |

## 4.2 Overlap

We were interested in any existing synergies between the three communities visible in the corpora. Thus, we determined all publications which were in more than one corpus (see figure 1). Notably, the intersections between the communities are rather small. At the same time this is only a syntactic overlap and does not allow any conclusions about references across corpuses or any overlap in the contents.

Nevertheless, there is no overlap between the exploratory programming corpus and the live coding corpus. The intersection of the live programming and the exploratory programming corpus is only two publications, which are both by authors of this paper. Finally, the intersection of the live coding and the live programming corpus contains seven publications.

Patrick Rein, Stefan Ramson, Jens Lincke, Robert Hirschfeld, and Tobias Pape

### 4.3 Venues and Journals

In order to give a first overview of the related communities, we determined venues and journals through which the single communities publish their results (see table 3). An interesting finding is that the publications on exploratory programming are spread across 21 collections, while the publications of live coding are spread across 24 collections although the corpus is three times larger. Further, live programming and exploratory programming share 7 collections, live programming and live coding share 5 collections, and exploratory programming and live coding only 2.

■ **Table 3** An overview over all venues and journals (denoted by *(J)*) from which publications are part of a corpus.

| Term (count of collections) | Venues / Journals, sorted alphabetically |
| --- | --- |
| live programming (30) | CHI, COP, ComposableWeb, Computer (J), ECOOP, FARM@ICFP, HCC, IBERAMIA, ICSE, International Journal of People-Oriented Programming (J), ISMM, Journal of Supercomputing Frontiers and Innovations (J), Journal of Visual Languages and Computing (J), LIVE, MOBILESoft, OOPSLA, Onward!, PLDI, PROMOTO, PVLDB, PX, Programming Journal (J), Science of Computer Programming (J), SIGMOD, SIGPLAN Notices (J), SPLASH, UIST, VISSOFT, VL/HCC, VRST |
| exploratory programming (21) | ACM SIGARCH Computer Architecture News (J), CHI, COMPCON, ECOOP, HOPL, ICSENG, IEEE Transactions on Software Engineering (J), IMCSIT, International Journal of Autonomic Computing (J), ISCA, Journal of Computer Based Instruction (J), Journal of Functional Programming (J), Journal of Visual Languages and Computing (J), Nordic Journal of Computing (J), OOPSLA, Onward!, PX, Software: Practice and Experience (J), User Modeling and User-Adapted Interaction (J), VL, VL/HCC |
| live coding (24) | AM, Blocks and Beyond, C&C, CLEI, Computer (J), Computers in Entertainment (J), Computer Music Journal (J), DUXU, EVA, EvoMUSART, FARM, ICFP, ICLC, ICMC, ICSE, Journal of Visual Languages and Computing (J), LIVE, MM, NIME, Organised Sound (J), OZCHI, SIGCSE, TEI, VL/HCC |

**Exploratory and Live, Programming and Coding**

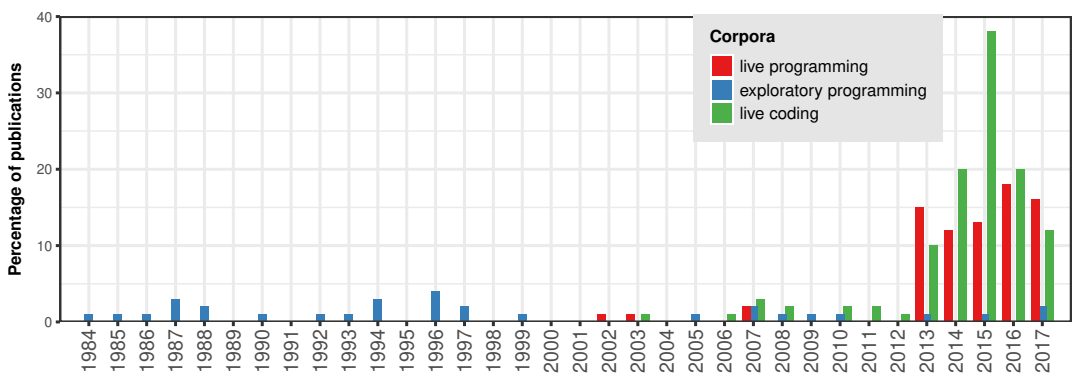

**Figure 2** A bar chart showing the chronological distribution of publications for each corpus. The height of the bar shows the absolute number of publications for that year.

## 5 Study Results

Besides basic bibliographic information, we also collected information specific for the individual research questions. In this section we report the data related to each question and discuss these results. For the results from tagging the corpora, we report the tags we used and the rules for applying them. Further, we report the inter-tagger reliability. Tagged data is shown in a normalized form as we are interested in comparing the distributions of tags between corpora (Detailed data can be found in appendix A).

### 5.1 Distribution over Time

From every publication we extracted the publication date (see figure 2). We show all historical distributions in absolute terms to illustrate the changing numbers of publications as an indicator of the changing size of the community.

Publications on the idea of *exploratory programming* (as defined for our sample) appear starting from 1984 ("Architecture of SOAR: Smalltalk on a RISC" [39]). At the same time, publications are evenly scattered over the past 30 years.
In contrast, in the *live programming* corpus the majority of work has been published within the past 15 years. This is interesting as Tanimoto brought the term *liveness* up in 1990 [36] but dedicated publications using the term *live programming* only began to appear in 2002 ("Toward a Unified Paradigm for Constructing and Understanding Robot Processes" [14]). Finally, in the *live coding* corpus the first publication is from 2003 ("Live Coding in Laptop Performance" [8]), which is also the most prominently cited publication on *live coding* within the corpus (see section 5.2).

### 5.2 Prominent Publications

In order to determine which publications define and form each research field, we determined the most prominent publications referenced by a corpus. Therefore, we collected all references from publications of each corpus. The cited publication did not have to be in the corpus itself in order to be collected. We list the five most cited references. If there are multiple publications with the same citation count at position five we included all publications with that citation count.

**Live Programming**
- (18) Real-time Programming and the Big Ideas of Computational Literacy [13]
- (17) VIVA: A Visual Language for Image Processing [36]
- (15) Smalltalk-80: The Language and Its Implementation [11]
- (13) It's Alive! Continuous Feedback in UI Programming [3]
- (13) Living It Up with a Live Programming Language [24]

**Exploratory Programming**
- (10) Smalltalk-80: The Language and its Implementation [11]
- (7) Smalltalk and Exploratory Programming [30]
- (5) Using Prototypical Objects to Implement Shared Behavior in Object-Oriented Systems [21]
- (4) An Efficient Implementation of SELF - A Dynamically-Typed Object-Oriented Language Based on Prototypes [6]
- (4) The Design and Implementation of the Self Compiler, an Optimizing Compiler for Object-Oriented Programming Languages [5]
- (4) Back to the Future: The Story of Squeak, a Practical Smalltalk Written in Itself [17]
- (4) Self: The Power of Simplicity [40]

**Live Coding**
- (33) Live Coding in Laptop Performance [8]
- (15) The Programming Language as a Musical Instrument [1]
- (14) Live Coding of Consequence [7]
- (14) Gibber: Live Coding Audio in the Browser [28]
- (13) Aa-Cell in Practice: An Approach to Musical Live Coding [32]
- (13) Live Algorithm Programming and a Temporary Organisation for its Promotion [41]

We retrieved less than a third of the references through Semantic Scholar. We extracted the remaining references in a semi-automated process directly from the publications.

The resulting publications show a clear orientation of the exploratory programming corpus towards Smalltalk and Self systems. The live programming corpus also includes one reference to Smalltalk but also refers to the thesis of Hancock which defines the concept of a *steady frame* [13] and Tanimoto's paper that defined the initial four levels of liveness [36]. The most prominent live coding publications do not share any publication with the other two corpora. The most prominent publication is at the same time the earliest publication in the corpus (see section 5.1).

**Exploratory and Live, Programming and Coding**

**Foundations of the Prominent Publications**   In order to get an impression of the foundations the communities refer to, we conducted a thematic analysis of the most prominent publications. In particular, we were looking for mentioned systems, types of systems, or research fields. Afterwards we aggregated the raw names into groups (for example, the Lisp systems groups includes "OakLisp" and the visual languages group includes "graphical languages"). We then selected all terms with 4 or more overall mentions (see table 4).

The great numbers of zero values in the exploratory programming column is a result of very few references being used in the corresponding publications. Several only refer back to Smalltalk and do not report on further related or previous work. Interestingly, live coding and live programming further both refer to "dynamic languages" and "visual languages". The overlap regarding spreadsheets is rather weak with only one mention in the live coding community [1].

Beyond the mere numbers, the mentions of visual languages are interesting as they all appear in the context of one of two general arguments. The first one is that visual languages have already supported liveness for a while but the transfer to textual languages was difficult (see for example [3]). The second argument is that visual languages do support liveness but only have a limited programming model (see for example [8]). Notably the references to visual programming languages in the live coding corpus often refer to graphical synthesizer programming environments.

■ **Table 4**   Number of publications per corpus which mention the system or field in the "Tags" column. The terms in the "Tags" column are representative for a number of similar tags (for example "Lisp systems" includes Interlisp-D).

| Tags | Live prog. | Exploratory prog. | Live coding | Total |
| --- | --- | --- | --- | --- |
| Lisp systems | 4 | 3 | 4 | 11 |
| Smalltalk | 3 | 5 | 2 | 10 |
| dynamic languages | 2 | 0 | 6 | 8 |
| visual languages | 3 | 0 | 3 | 6 |
| SuperCollider | 0 | 0 | 5 | 5 |
| spreadsheets | 4 | 0 | 1 | 5 |
| ChucK | 0 | 0 | 4 | 4 |
| Forth | 0 | 0 | 4 | 4 |
| data-flow | 4 | 0 | 0 | 4 |

## 5.3 Application Domains

In order to find out in which contexts *liveness* is applied in, we determined the intended application domains of the live aspect described in publications (see figure 3). We only tagged domains which were major themes of the work and were not only mentioned as an example. If no particular domain was given most works were tagged as *general software development*. We only assigned a *can not be determined* tag if the work was about some aspect of liveness which can not be tied to any particular practical activity in any domain. We only assigned multiple tags to publications whenever the focus was actually on two domains.

**Artificial intelligence**  The design and implementation of artificial intelligence systems.

**Audio performance art**  The in-time creation of audible art in front of an audience or for recording.

**Computer games**  The design and implementation of computer games, excluding game engines.

**Computer graphics**  Creating programs concerned with generating visual output (though not as a performance in contrast to visual performance art)

**Data analysis**  The analysis of existing data to gain insights.

**Databases**  The programming and manipulation of databases of various kinds.

**Education and training**  The program is created to either learn programming or to learn something from the program (for example a model of ecological processes used as a demonstration tool in teaching).

**Enterprise-Resource-Planning (ERP) systems**  The design and implementation of ERP systems

**General software development**  General software development with no stated focus on a particular application domain.

**Performance art**  Any form of performance other than pure visuals and audio, for example writing poems, dance, general performance art reflections.

**Physical computing**  All systems which have some interface with the physical world (actuators or sensors).

**Programming languages**  The design and implementation of programming languages (syntax and semantics).

**Simulation**  Systems and programs for simulation of the physical world.

**Software verification**  The verification of a program, either through a tool or through writing a specification to be run against a program

**Systems programming**  Development of execution environments or fundamental libraries

**Text editing**  Creating and editing text (possibly rich text)

**User interfaces**  The design and implementation of graphical user interfaces.

**Virtual/Augmented reality**  The design and implementation of virtual or augmented reality systems

**Visual performance art**  The in-time creation of visual art in front of an audience or for recording.

**Web development**  The construction of applications running in or on the web. This does not involve the construction of programs through an environment implemented as a web application.

### 5.3.1 Results

The most significant result is that the majority of publications on exploratory programming and live programming have no particular application domain but deal

**Exploratory and Live, Programming and Coding**

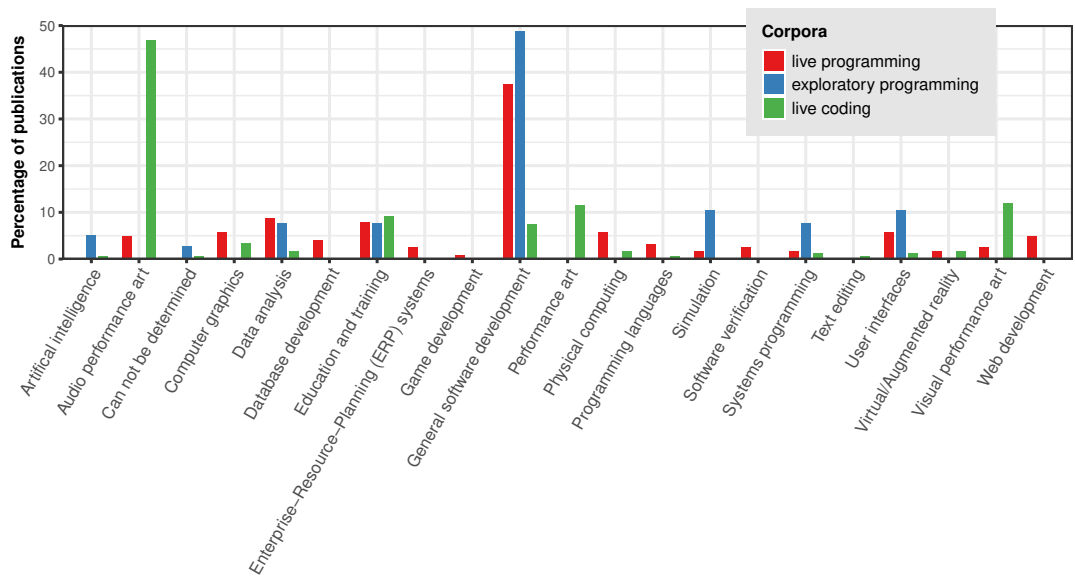

▪ **Figure 3** A bar chart showing the relative distribution of application domains referred to in the publications. The height of the bar indicates the percentage of publications having that tag within the respective corpus.

with general software development. Besides the main domain of each corpus, the next major domains for exploratory programming are simulation and user interface programming. For live programming the next major domains are education and data analysis. In contrast, as expected, the live coding publications focus very much on performance art, in particular audio performance art. Taking all performance art categories together, more than half of the publications on live coding are concerned with applying liveness in performance art. When taking all performance art domains as one domain the following two domains for live coding are education and training and general software development.

An interesting aspect is the fact that proportion of papers on education and training is about the same in all three corpora.

The inter-tagger reliability score for this dimension is 0.69, which indicates good agreement.

### 5.4 Motivations for Liveness

While liveness is generally an intriguing idea, the motivations for it vary greatly between authors and communities (see figure 4). However, the initial motivation influences designs and research methodology and is thus an integral element of the definition of a field.

The following tags reflect the variety of actual motivations for liveness. When determining the motivation in a publication we only assigned a tag when it was a) explicitly stated or b) a central theme of the work which might be stated in other words and not in relation to liveness directly. As we extracted the list directly from the publications, the items do not have the same level of abstraction. Again, these tags are not mutually exclusive.

**Accessibility** Liveness makes programming more accessible. This involves comprehension but is more specific towards enabling less experienced programmers

**Can not be determined** We added this code explicitly in order to determine how often liveness is taken as granted.

**Collaboration** The liveness enables collaboration between users as they can see each other"s changes

**Comprehension** Ease the process of comprehension and learning of program behavior or the program domain

**Creativity** This applies to all works explicitly mentioning creativity as their primary motivation.

**Engagement** Getting users more engaged through liveness

**Estrangement** Live programming as an extreme form of programming which enables reflections on the nature of programming

**Exploration** This is the very specific notation of liveness enabling programmers to explore existing systems, APIs, or future designs (It does not suffice to say it supports *exploratory programming* as this would in most cases be self-referential; mentioning *exploratory work* falls in this category though).

**In-time control** Liveness is an integral part of the activity itself. The output has to be created and modified with a short delay as the changing of the output is the relevant action; this does often correlate with live coding but does not have to.

**Live tuning** Liveness as a means to explore a continuous parameter space of a design

**Productivity** Liveness improves productivity, most often through speeding up some activity during programming

### 5.4.1 Results

As the term exploratory programming suggests, the major motivation in the corresponding corpus is *exploration*. However, almost a third of exploratory programming publications did not state a motivation for liveness in their publications. In contrast, the motivation for liveness in the live coding corpus is the in-time control of some output (this matches well with the application domain focus being performance art as shown in figure 3). The second motivation for liveness in live coding is collaboration which is almost not covered by the other corpora at all. The major motivations for liveness in the live programming corpus are accessibility and comprehension.

Productivity, although still somewhat relevant to all three corpora is not a dominant motivation for any corpus.

The inter-tagger agreement for this dimension was 0.44 which indicates fair agreement. We investigated the source for this disagreement and refined the definitions for the tags. Further, we determined that the disagreement stemmed from the fact that the first tagger interpreted major themes more strictly and thereby assigned less tags overall.

**Exploratory and Live, Programming and Coding**

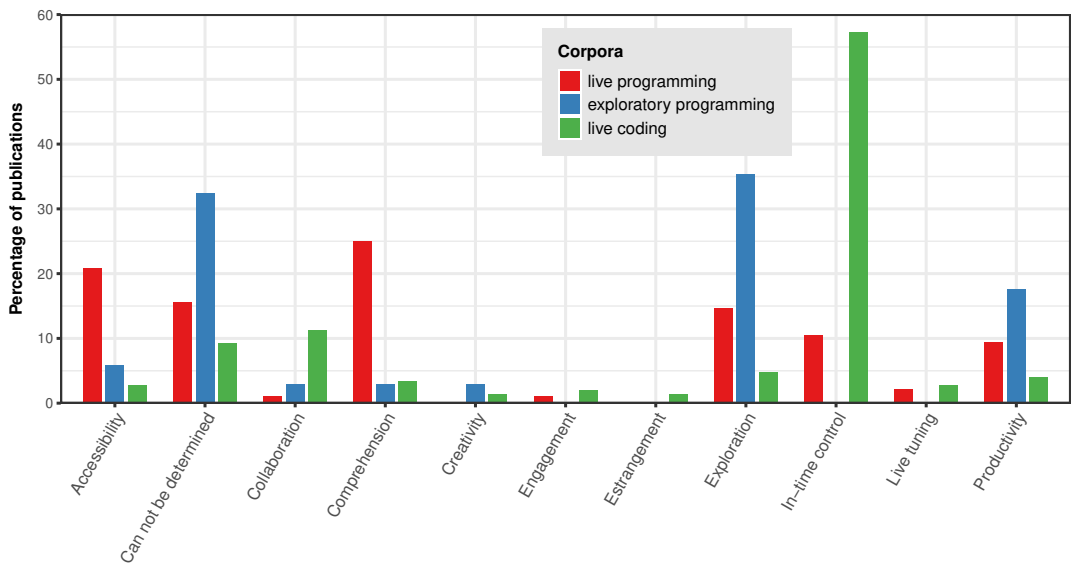

**Figure 4** A bar chart showing the relative distribution of motivations for liveness as given in the publications. The height of the bar indicates the percentage of publications having that tag within the respective corpus.

## 5.5 Types of Contributions

As illustrated before, the three communities differ greatly in their motivations for liveness and the desired outcomes. Consequently, the communities apply different methodologies and value different kinds of contributions which advance their field.

We tagged the contribution of each publication as described. We did not tag any contributions implied by the larger context. For example, when the design of a new interaction within an existing tool is described in a publication, the interaction is the contribution not the tool. Some of these tags are hierarchical, so a tag can have several sub-tags which also count as being part of the super-tag set (see the empirical tags).

**Empirical studies** Works which empirically evaluated an artifact or intervention. As this covers a wide range of contributions we made codes more specific with the following additional information: a) fixed / flexible setup[2] b) quantitative / qualitative outcome c) humans as subject (for example usability study) / systems as subjects (for example performance benchmarks) [29]

**Empirical studies on liveness** While the previous category covers evaluations of concrete artifacts and interventions, studies in this category try to uncover the impacts of liveness. We used the same sub-tags as the previous category.

**Survey** Surveys of existing artifacts, for example publications or systems. These can count as empirical contributions whenever some form of planning of the survey

---

[2] The methodology was completely determined before the observation (fixed); The methodology was adapted during the observation (flexible) [29].

was described. Non-empirical surveys give an overview of the field without any formal methodology (neither fixed nor flexible)

**Technical design** The description of a technical implementation including implementation details.

**User interaction design** Describes the design of a particular user interaction on several levels. This includes input devices and mechanisms, feedback mechanisms, as well as workflows.

**Programming interface design** This involves programming language design (syntax and semantics) as well as the design of the interfaces of libraries (APIs).

**Programming tool design** The description of the design of a programming tool, workbench, or environment [10]

**Individual account of experience** An individual subjective account which was not planned as an empirical study

**Formal methods** The description of a formal model or the application of an existing model

**Algorithm design** The description of an algorithm without providing details about its implementation. This can be presented for example through a textual description, using pseudo code, or a mathematical model.

**Practices** The description of practices or workflows which were either empirically observed or from subjective experience

**Design principles** A collection of rules or values for designs

**Essay** An application of theoretical models or principles to an existing phenomena. This definition is intentionally very general. Essays are any form of structured argumentation which is not based on empirical evidence or a novel design.

**Performance** The description of an artistic performance

**Curriculum design** The description of a new curriculum or teaching model

#### 5.5.1 Results

The first significant result that there is no strongly significant type of outcome in the live coding corpus. Actually, the live coding corpus covers all types of contributions. Further, all three corpora have their major focus on tool designs and technical designs. Only the live programming corpus differs slightly as the publications primarily focus on the design of tools and only second on technical designs. Again the focus of the live coding community on performance art might explain the high percentage of essays and descriptions of performances and practices.

As the empirical studies taken together are a major part of the contributions of each corpus, we examined them in greater detail which can be seen in figure 6. The overall distribution is very similar for all three corpora. The only major difference is that the live coding publications have a stronger emphasis on flexible qualitative studies with users. At the same time they deemphasize fixed quantitative studies on systems in comparison with the other two corpora.

The inter-tagger agreement for this dimension is 0.59 which indicates good agreement.

**Exploratory and Live, Programming and Coding**

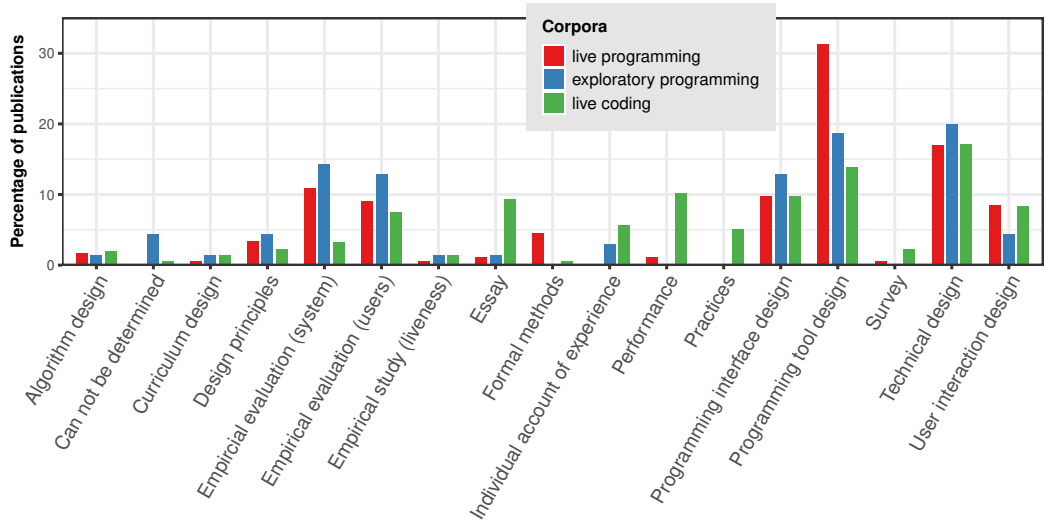

**Figure 5** A bar chart showing the relative distribution of types of contributions described in the publications. The height of the bar indicates the percentage of publications having that tag within the respective corpus.

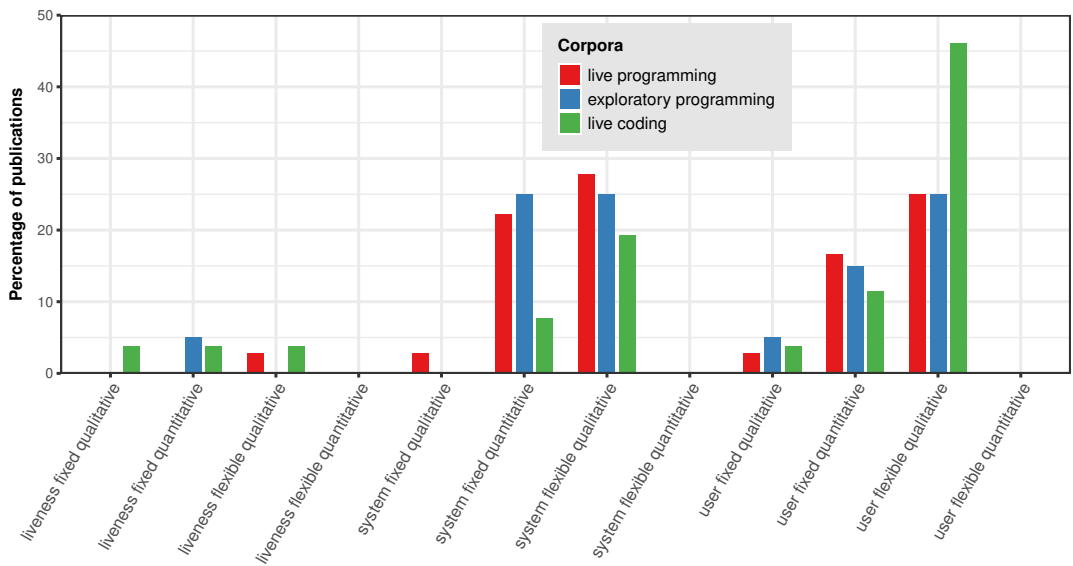

**Figure 6** A bar chart showing the relative distribution of types of empirical studies described in the publications. The height of the bar indicates the percentage of publications having that tag within the respective corpus.

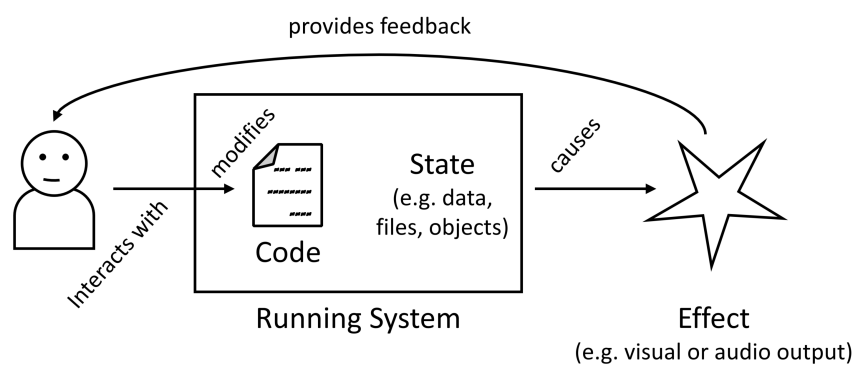

**Figure 7** An illustration of the relation between evolving a running system, modifying code and creating a program, and causing an effect. Programmers do all three throughout the programming process but aim for one of the three as the outcome.

## 5.6 Intended Outcome

The notion of *an impression of changing a program while it is running* subsumes different intended outcomes of *running* a system. For example, users of Smalltalk systems experience liveness as changing a running system, while a performance artist coding aims to create an effect in the real world. Notably, both programmers interact with a running system and modify meta-structures of a program (see figure 7). However, the ultimately intended outcome differs between the two.

We assigned three intended outcomes from *running a system* according to the intended outcome. As all three outcomes for running a system are possible for most activities and programs (for example a programmer might keep a Python REPL for data analysis open for a long time). Therefore, we assigned the code only to the emphasis in the described work. Although, the categories are mutually exclusive, we allowed assigning multiple tags to a publication when it described multiple aspects of the activity of programmers.

**Causing an effect** The programmer wants some output and therefore runs a program. The program is only secondary and might be discarded after the computation, this covers in-time output such as generated sound as well as discrete outputs created by a data analysis algorithm.

**Creating a program** The programmer wants to create a program. The program is currently running to get feedback on the dynamic behavior of it through side effects. The resulting artifact, however, is some static representation of the program behavior.

**Evolving a system** The programmer wants to evolve a system which is constantly running. Thus, programmers change the representation of the program in the system as well as the state of the system (often the representation is part of the state). Examples are image-based systems or database systems.

**Not applicable** For some publications the intended output can not be determined which is represented by this tag.

**Exploratory and Live, Programming and Coding**

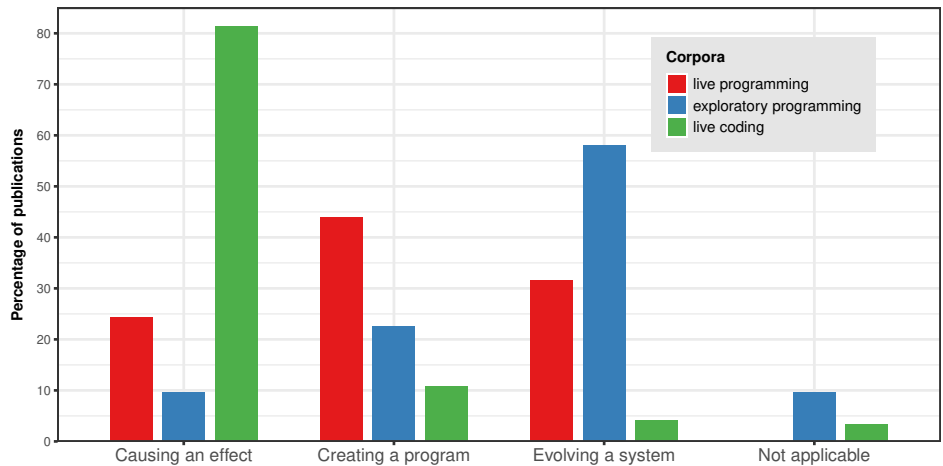

**Figure 8** A bar chart showing the relative distribution of intended outcomes. The height of the bar indicates the percentage of publications having that tag within the respective corpus.

#### 5.6.1 Results

The results for this tagging show a clear tendency for each corpus. The exploratory programming corpus focuses on evolving a system while the live coding corpus focuses on causing an effect. The live programming publications focus on creating a program, however, the focus is not as strong as it is with the other two corpora.

The inter-tagger agreement for this dimension is 0.70 which indicates good agreement.

### 5.7 Keywords

In order to determine other perspectives on the field of research on programming related to liveness, we extracted the keywords of each publication (see table 5). We only recorded keywords mentioned in the actual publication. We also only recorded the keywords chosen by the authors and ignored general tags and categories predetermined by the publisher (for example the ACM subject classification).

We cleaned the data marginally by merging the keywords "programming environment" and "programming environments" as well as "end-user development" and "end-user programming". Finally, we selected all keywords mentioned more than once.

An interesting aspect is the overlap between the keywords. The live programming keywords include live coding and the exploratory programming keywords include live programming. The live coding keywords contain neither of the other two terms. Other keywords shared between at least corpora are direct manipulation, programming environment, liveness, and music.

**Table 5** An overview over all keywords used in publications in the corpora.

| Term (works with terms / corpora size) | Keywords (count), sorted by count |
|---|---|
| live programming (53/78) | live programming (33), live coding (6), end-user programming (5), debugging (4), direct manipulation (4), Smalltalk (4), live programming environment (3), music (3), mashups (3), spreadsheets (3), programming environment (3), Datalog (3), web services (2), prototyping (2), integrated development environment (2), data analysis (2), incremental maintenance (2), streaming data (2), LogiQl (2), testing (2), LogicBlox (2), Javascript (2), liveness (2), mashup tools (2), data mining (2) |
| exploratory programming (14/31) | exploratory programming (6), programming environment (4), virtual machine (2), gestures (2), programming by demonstration (2), live programming (2), Forms/3 (2), direct manipulation (2), liveness (2), Self (2) |
| live coding (34/112) | live coding (25), music (6), generative art (4), visualization (2), software engineering (2), visual programming (2), liveness (2), art (2), domain specific languages (2), creative coding (2), generative music (2), programming environment (2), functional programming (2), music performance (2), web audio (2), improvisation (2) |

## 6 Discussion

Taking into account the results for each research question, we discuss the implications for the relation between the three communities. From the results we can determine an overall focus of each corpus and thereby the corresponding community. First, we focus on each particular community. Then we examine how these communities differ in the ends they want to achieve through liveness. Based on these ends we look at the particular contributions of the single communities and how they might be combined. In contrast to these potential synergies we also point out future work for all communities and potential new application domains. Finally, we discuss threats to the validity of the results, as well as issues arising when generalizing the findings.

**Exploratory and Live, Programming and Coding**

**6.1 Summary: Focus of each Community**

The major types of contributions in all three corpora are the design of tools or technical artifacts and corresponding evaluations of some kind. Thus, the three corpora hint that all three communities mainly engage in design science [19].

According to our results, the idea of *live programming* is mostly motivated by improving accessibility to and comprehension for programming. The corpus hints that the community contributes to this goal by designing programming tools which are evaluated through user studies. The major intended outcome for interacting with a running system is to create a program.

The *live coding* corpus focuses on performance art, as expected. The publications advance the concept with a broad variety of contributions. When working empirically, these publications tend to provide insights into the qualitative aspects of live coding. Although the community has only formed rather recent, there seems to be a strong tradition linking the publications, as can be seen in the prominent publications which are very much focused on original publications on live coding. The intended outcome of interacting with a running system is clearly to create an effect.

The idea of *exploratory programming* is motivated by enabling users to explore existing systems or new ideas and designs. The corpus hints that the community values technical designs for enabling exploration. Further, other major contributions are tool designs and empirical studies with users and systems. Further, the keywords and the most prominent publications suggest that a lot of work in the exploratory corpus is about Smalltalk and SELF systems.[3] There are less hints for a focus on few systems in the live programming and live coding corpora.

**6.2 The Same Means for Different Ends**

Common to work on liveness in all three communities is a focus on "immediate feedback". Crucial to all approaches is a short feedback loop between changing some symbolic representation of a program and getting some meaningful output related to that change, independent of whether this feedback has to be fetched manually or is provided automatically. Further, they also share a common technological root which is dynamic languages, in particular Lisp, and to some degree Smalltalk. Beyond these commonalities, when looking at the original motivations for liveness and the foundations of their most prominent publications, we can see that the three communities want to make use of the means of "liveness" for different ends.

The live coding community aims for short round-trip times out of necessity. In order to create music live in front of an audience the artist has to be able to create an impression of controlling the produced effects in real-time (or close to it). Some performers have also used compiled languages which reportedly resulted in silence for several seconds during performances which was sometimes deemed an aesthetic issue [8, 20]. Further, several live coding publications refer to the general field of automatic composition of music and the earlier practice of live patching, which is the live configuration of virtual or real synthesizers. In this context, live coding is a turn towards a more expressive interface as described in one of the earliest publications on live coding: "those suspicious of the fixed interfaces and design decisions of such software turn to the customizable computer language." [8]

---

[3] This should not be taken as a statement about the publications on *exploratory programming* but on *exploratory programming environments*.

In contrast to the necessity which drives liveness in live coding, an exploratory workflow does not require liveness. At the same time it can greatly benefit from it as described in one of the corresponding, prominently cited papers: "To have exploratory programming be successful, the cost of experimentation must be low. The time to write the code for an experiment must be short enough that the code can be discarded if the idea fails to produce the desired result." [30] Trying out different ideas is easier when the time between having an idea and getting feedback on it is short.

Finally, the live programming tools want to explicitly leverage liveness in order to make the dynamic behavior of ordinary programs or parts of programs easier to understand [13]. Many works are about making the dynamic behavior accessible by making it visible and explorable. Correspondingly, the fundamental works on live programming refer to fields also dealing with similar issues such as spreadsheets, visual programming, and data-flow execution models [3].

### 6.3 Potential for Learning From Each Other

The live coding community contributes to the overall field actual insights into the practices of programming in a live system. The live coding community draws methods from arts and humanities to investigate the individual experiences with live systems. The results includes reports on individual practices, particular performances, and individual perspectives of what makes up the art of live coding. In contrast, the live programming and exploratory programming community mostly either describe idealized workflows or abstract results from usability studies. In order to gain deeper insights and inspirations for future designs the other two communities could build upon the live coding work.

On a similar note, the live coding community can provide insights into programming as a social phenomena as well as designs for supporting collaboration for creating programs with short feedback loops.

Beyond that the major potential for synergies could emerge from the different motivations and intended outcomes. The live programming community provides a variety of tool designs and implementation strategies to provide immediate feedback for particular use cases or parts of systems for example user interfaces or data analysis algorithms. Some of these designs and interfaces might offer new dimensions to be used as compositional elements in live coding. At the same time the live programming designs are often rather specialized and do for the most part not provide a general experience or concept of liveness beyond their use case or scenario.

The perspectives of exploratory programming environments (mainly based on Smalltalk- and LISP-derived systems) could be beneficial here by providing a more wholesome approach. These systems are built upon a fundamental notion of liveness

**Exploratory and Live, Programming and Coding**

which in some instances is even available to programmers when dealing with faults in the base system. At the same time these wholesome systems often do not provide specialized means to shorten the feedback loop for example when writing a data analysis algorithm. In these cases, the exploratory environments could benefit from the specialized tool designs created by the live programming community.

### 6.4 A Common Future Work: Empirical Evidence

There is another commonality between the three communities which is the lack of empirical evidence of many of the claims. While the live coders might not have to justify the liveness as it is inherent to their activity, the exploratory and live programming communities should at one point be able to prove their claims of "liveness" improving comprehension and exploration. Most empirical work in the communities is done on evaluating particular designs or interventions. Only a few studies investigate the particular effects of liveness [18, 16, 25, 34, 15]. Besides these studies on the effects of liveness within the corpus, there is one other prominent study outside the corpus conducted in the realm of visual languages [42]. These few studies are not enough yet to reach any conclusion what the beneficial and adverse effects of liveness are and leave much room for future work in all three communities investigated in this study, and potentially also for adjacent communities.

### 6.5 Future Application Domains

Looking at the application domains covered by each community, there is still a lot of potential for beneficial combinations. For example, there is only a single work in the domain of scientific simulations which aims at creating an effect. While interactive simulations have often been proposed as a medium for education, and live coding does cover the live creation of visual simulations, there seems to be little work on programming interactive ones live in talks, presentations, or even performances [9]. Another surprisingly uncovered domain in the corpora is game development. In particular, there are no works on the intersection between liveness for evolving a running system and game development. This is surprising as some game engines and their corresponding tools do already seem to provide a development experience similar to liveness for evolving a running system. Recent work has described some of the implementation strategies but a thorough coverage still seems to be missing [33].

Further, the communities have only marginally covered some application domains, such as machine learning, distributed applications and protocols, computer security, model-driven software development, and infrastructure administration. Some of these topics might not seem like they could benefit from liveness. However, works of the live programming community on liveness in software verification have shown that even for quite formal application domains promising tool designs can emerge [26]. Further, communities around these application domains might very well have tools and environments providing a live experience, however the examined communities have not taken note of it yet.

## 6.6 Threats to Validity

We identified threats to the validity to the individual results based on our search process, the available data at indexing services, and potential tagger bias. Further, there are also issues related to the degree to which we can generalize our findings because of our selection of initial terms and because we did not include the publications cited by the publications in the original corpora.

### 6.6.1 Validity of Results

The major threat to the validity of the results is our search process which was based on three major indexing services. This limits our results to indexed venues and journals. For example, this excluded the publications from the meetings of the Psychology in Programming Interest Group (PPIG). Further, by focusing on peer-reviewed publications we also excluded all work done in a less formal setting which might be substantial depending on the community. For example, works published through the TOPLAP web page [37] do not appear in our corpus as well as the articles and demonstrations of Bret Victor which the live programming and the live coding community often reference as inspiration.[4] Consequently, the results have to be seen as a description of the academic perspective on the three ideas. Further, the data provided by the indexing services might be incomplete or wrong in relation to the actual content or bibliographic information of the publications.

Regarding the comparison of the distributions of the three corpora it is important to be aware of the small sample size for exploratory programming. If some tag was assigned to 5 % of the exploratory programming publications that means that we assigned that tag to three publications. Hence, the exploratory programming corpus is more sensitive to, for example, accidentally missed out tags.

Similarly, the exploration of the foundations of the individual fields is currently based on the most prominent publications only. This is a very small sample and an extensive investigation of the historic developments from which these communities emerged is promising future work.

The inter-tagger agreement for the data on motivation is low and, depending on the Kappa index used, only indicates fair agreement. When looking at the reason for the stark disagreement we determined that the first tagger was more conservative in assigning motivations which were not explicitly stated. However, this flexibility in the synthesis of this dimension might have introduced bias.

### 6.6.2 Generalization of Results

The missing hull of cited publications is the major issue regarding the generalization of our findings from insights into publications on an idea to insights into the values and methods of a community. Notably, several of the most prominent publications of each corpus are not themselves part of the corpus (see section 5.2). These cited papers might include valuable insights for the values and methods of each community although they do not explicitly state the term. While our current search and appraisal process keeps the focus of each corpus on one term it also excludes the overall set which might be relevant to a community. Another consequence of the focus on the initially chosen terms *live programming*, *live coding*, and *exploratory programming* is that the corpora do not cover the whole spectrum of work on liveness. Several significant publications are left out, such as the paper on the Morphic system, the original VIVA paper introducing the first four levels of liveness, or later works extending on these levels [22, 36, 4].

---

[4] For example, the article "Learnable Programming" (http://worrydream.com/#!/LearnableProgramming (accessed 14th of May 2018))

Also, the selection of these three terms very much represents the *live programming* perspective and the associated terms. Although the analysis of keywords did not yield another similar term, we are aware of terms such as *interactive programming* or *in-time programming* also being used to describe similar ideas. The significance and nature of these concepts might be promising future work.

Finally, the exploratory programming term yielded significant less publications than the other two terms. This is an issue as it was chosen to cover the perspective of Lisp and Smalltalk systems, which according to the prominent publications, has a significant influence on the communities. Further, the exploratory programming corpus consists mostly of works on particular Smalltalk and Lisp systems. This in turn means that research on similar systems, for example reactive databases, might not be covered.

# 7 Conclusion

The idea of *liveness* as creating an *impression of changing a program while it is running* has been discussed in academic communities from the perspective of three different concepts: *live programming*, *exploratory programming*, and *live coding*. Although all three contain this idea as an important part, there is only little overlap in publications or venues. Further, important contributions to the idea of liveness are spread over all three concepts and might not be easily accessible to researchers new to the field. In this paper, we conducted a study on a sample of publications on the three concepts to gain insights into their particular focus and potential overlaps.

In summary, all three communities conduct design science as they create tool designs, describe technical designs, and evaluate them in empirical studies. At the same time, they differ in their intended outcomes, and motivations for liveness. Work on exploratory programming is mostly motivated by enabling exploration during the evolution a running system. Research on live coding is motivated by controlling in-time output while supporting programmers in creating an effect in the real world through computation. Finally, live programming is motivated by making the activity of creating a program more comprehensible and accessible.

The determined differences and commonalities might give rise to interesting future work when combining contributions from all three fields, given that researchers will take interest in the other fields.

**Acknowledgements** We gratefully acknowledge the financial support of the Research School for Service-oriented Systems Engineering of the Hasso Plattner Institute and the Hasso Plattner Design Thinking Research Program.

[10] Alfonso Fuggetta. "A Classification of CASE Technology". In: *Computer* 26.12 (1993), pages 25–38. ISSN: 0018-9162. DOI: 10.1109/2.247645.

[11] Adele Goldberg and David Robson. *Smalltalk-80: The Language and Its Implementation*. Boston, MA, USA: Addison-Wesley Longman Publishing Co., Inc., 1983. ISBN: 0-201-11371-6.

[12] Maria Grant and Andrew Booth. "A Typology of Reviews: An Analysis of 14 Review Types and Associated Methodologies". In: *Health Information & Libraries Journal* 26.2 (2009), pages 91–108. DOI: 10.1111/j.1471-1842.2009.00848.x.

[13] Christopher Hancock. "Real-Time Programming and the Big Ideas of Computational Literacy". PhD thesis. Massachusetts Institute of Technology, 2003.

[14] Christopher Hancock. "Toward a Unified Paradigm for Constructing and Understanding Robot Processes". In: *Proceedings of the IEEE Symposia on Human Centric Computing Languages and Environments (HCC) 2002*. HCC '02. Washington, DC, USA: IEEE Computer Society, 2002, pages 107–109. ISBN: 0-7695-1644-0. DOI: 10.1109/HCC.2002.1046362.

[15] Christopher Hundhausen and Jonathan Brown. "An Experimental Study of the Impact of Visual Semantic Feedback on Novice Programming". In: *Journal of Visual Languages and Computing* 18.6 (2007), pages 537–559. ISSN: 1045-926X. DOI: 10.1016/j.jvlc.2006.09.001.

[16] Christopher Hundhausen and Jonathan Brown. "What You See Is What You Code: A "Live" Algorithm Development and Visualization Environment for Novice Learners". In: *Journal of Visual Languages and Computing* 18.1 (2007), pages 22–47. ISSN: 1045-926X. DOI: 10.1016/j.jvlc.2006.03.002.

[17] Daniel Ingalls, Ted Kaehler, John Maloney, Scott Wallace, and Alan Kay. "Back to the Future: The Story of Squeak, a Practical Smalltalk Written in Itself". In: *Proceedings of the Conference on Object-Oriented Programming Systems, Languages & Applications (OOPSLA) 1997*. Volume 32. 10. Atlanta, Georgia, USA: ACM, 1997, pages 318–326. DOI: 10.1145/263698.263754.

[18] Jan-Peter Kramer, Joachim Kurz, Thorsten Karrer, and Jan Borchers. "How Live Coding Affects Developers' Coding Behavior". In: *Proceedings of the Symposium on Visual Languages and Human-Centric Computing (VL/HCC) 2014*. Melbourne, VIC, Australia: IEEE Computer Society, 2014, pages 5–8. DOI: 10.1109/VLHCC.2014.6883013.

[19] Bill Kuechler and Vijay Vaishnavi. "On Theory Development in Design Science Research: Anatomy of a Research Project". In: *European Journal of Information Systems* 17.5 (2008), pages 489–504. DOI: 10.1057/ejis.2008.40.

[20] Sang Lee and Georg Essl. "Communication, Control, and State Sharing in Networked Collaborative Live Coding". In: *Ann Arbor* 1001.48109 (2014), pages 2121–2121. URL: http://www.nime.org/proceedings/2014/nime2014_554.pdf.

## A  Detailed Data Sets

The following tables are the absolute numbers behind the results shown in section 5. Note that again the rows do not have to add up to the total column as publications can be in more than one corpora. Further, the total row does not add up to the total number of papers per corpus as tags can be assigned multiple times.

### A.1 Application Domains

| Tag | All | Live prog. | Expl. prog. | Live coding |
| --- | --- | --- | --- | --- |
| Artificial intelligence | 3 | 0 | 2 | 1 |
| Audio performance art | 84 | 6 | 0 | 82 |
| Computer graphics | 11 | 7 | 0 | 6 |
| Data analysis | 15 | 11 | 3 | 3 |
| Database development | 5 | 5 | 0 | 0 |
| Education and training | 25 | 10 | 3 | 16 |
| ERP systems | 3 | 3 | 0 | 0 |
| Game development | 1 | 1 | 0 | 0 |
| General software development | 74 | 47 | 19 | 13 |
| Performance art | 20 | 0 | 0 | 20 |
| Physical computing | 10 | 7 | 0 | 3 |
| Programming languages | 5 | 4 | 0 | 1 |
| Simulation | 6 | 2 | 4 | 0 |
| Software verification | 3 | 3 | 0 | 0 |
| Systems programming | 7 | 2 | 3 | 2 |
| Text editing | 1 | 0 | 0 | 1 |
| User interfaces | 13 | 7 | 4 | 2 |
| Virtual/Augmented reality | 5 | 2 | 0 | 3 |
| Visual performance art | 22 | 3 | 0 | 21 |
| Web development | 6 | 6 | 0 | 0 |
| Can not be determined | 2 | 0 | 1 | 1 |
| Total | 321 | 126 | 39 | 175 |

### A.2 Intended Outcome

| Tag | All | Live prog. | Expl. prog. | Live coding |
| --- | --- | --- | --- | --- |
| Causing an effect | 114 | 20 | 3 | 97 |
| Creating a program | 53 | 36 | 7 | 13 |
| Evolving a system | 48 | 26 | 18 | 5 |
| Not applicable | 7 | 0 | 3 | 4 |
| Total | 222 | 82 | 31 | 119 |

# Exploratory and Live, Programming and Coding

## A.3 Types of Contributions

| Tag | All | Live prog. | Expl. prog. | Live coding |
| --- | --- | --- | --- | --- |
| Algorithm design | 8 | 3 | 1 | 4 |
| Curriculum design | 4 | 1 | 1 | 3 |
| Design principles | 13 | 6 | 3 | 5 |
| Empircial evaluation (system) | 36 | 19 | 10 | 7 |
| Empirical evaluation (users) | 34 | 16 | 9 | 16 |
| Empirical study (liveness) | 5 | 1 | 1 | 3 |
| Essay | 23 | 2 | 1 | 20 |
| Formal methods | 9 | 8 | 0 | 1 |
| Individual account of experience | 14 | 0 | 2 | 12 |
| Performance | 22 | 2 | 0 | 22 |
| Practices | 11 | 0 | 0 | 11 |
| Programming interface design | 44 | 17 | 9 | 21 |
| Programming tool design | 94 | 55 | 13 | 30 |
| Survey | 5 | 1 | 0 | 5 |
| Technical design | 78 | 30 | 14 | 37 |
| User interaction design | 34 | 15 | 3 | 18 |
| Can not be determined | 4 | 0 | 3 | 1 |
| Total | 438 | 176 | 70 | 216 |

## A.4 Motivations

| Tag | All | Live prog. | Expl. prog. | Live coding |
| --- | --- | --- | --- | --- |
| Accessibility | 25 | 20 | 2 | 4 |
| Collaboration | 19 | 1 | 1 | 17 |
| Comprehension | 27 | 24 | 1 | 5 |
| Creativity | 3 | 0 | 1 | 2 |
| Engagement | 4 | 1 | 0 | 3 |
| Estrangement | 2 | 0 | 0 | 2 |
| Exploration | 32 | 14 | 12 | 7 |
| In-time control | 91 | 10 | 0 | 86 |
| Live tuning | 6 | 2 | 0 | 4 |
| Productivity | 18 | 9 | 6 | 6 |
| Can not be determined | 40 | 15 | 11 | 14 |
| Total | 267 | 96 | 34 | 150 |

## About the authors

**Patrick Rein** is a PhD student in the Software Architecture Group of the Hasso Plattner Institute at the University of Potsdam. His research interests include live and exploratory programming systems as well as personal information management systems. Contact Patrick at patrick.rein@hpi.uni-potsdam.de.

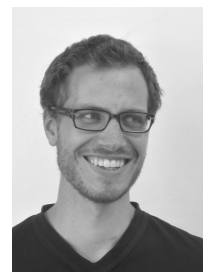

**Stefan Ramson** is a doctoral researcher at the Software Architecture Group. His research interests include programming language design and natural programming. Contact Stefan at stefan.ramson@hpi.uni-potsdam.de.

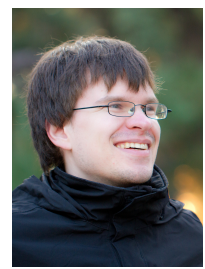

**Jens Lincke** is a member of the Hasso Plattner Institute's Software Architecture Group. His research interests include live and explorative programming. Lincke received a PhD in IT-Systems Engineering from the Hasso Plattner Institute at the University of Potsdam. Contact Jens at jens.lincke@hpi.uni-potsdam.de.

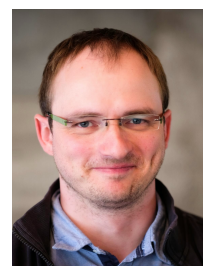

**Robert Hirschfeld** leads the Software Architecture Group of the Hasso Plattner Institute at the University of Potsdam. His research interests include dynamic programming languages, development tools, and runtime environments to make live, exploratory programming more approachable. Hirschfeld received a PhD in computer science from Technische Universität Ilmenau. Contact Robert at robert.hirschfeld@hpi.uni-potsdam.de.

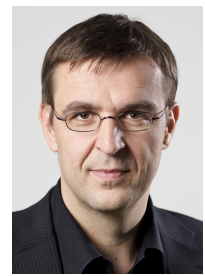

**Tobias Pape** is a PhD student in the Software Architecture Group of the Hasso Plattner Institute at the University of Potsdam. He is interested in virtual-machine construction, language design, and data structure optimization. Contact Tobias at tobias.pape@hpi.uni-potsdam.de.

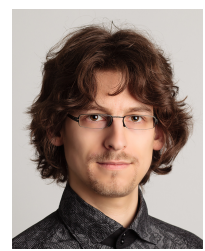